\documentclass[12pt]{article}
\usepackage{times}
\usepackage[letterpaper, portrait, margin=1in]{geometry}
\usepackage{useltp_ksp_1.2}
\usepackage[utf8]{inputenc}
\usepackage{enumitem,amssymb}
\usepackage{graphicx}
\newlist{thematic}{itemize}{8}
\setlist[thematic]{label=$\square$}
\usepackage{pifont}
\usepackage{setspace,wrapfig}
\usepackage{multicol}
\usepackage[sort]{natbib}
\usepackage[compact]{titlesec}
\newcommand{\cmark}{\ding{51}}%
\newcommand{\done}{\rlap{$\square$}{\raisebox{2pt}{\large\hspace{1pt}\cmark}}%
\hspace{-2.5pt}}

\let\OLDthebibliography\thebibliography
\renewcommand\thebibliography[1]{
  \OLDthebibliography{#1}
  \setlength{\parskip}{0pt}
  \setlength{\itemsep}{0pt plus 0.3ex}
}

\begin{document}

\huge
Characterizing the Atmospheres of Irradiated Exoplanets at High Spectral Resolution

\normalsize

\noindent \textbf{Thematic Areas:} \hspace*{60pt} \done Planetary Systems \hspace*{10pt} $\square$ Star and Planet Formation \hspace*{20pt}\linebreak
$\square$ Formation and Evolution of Compact Objects \hspace*{31pt} $\square$ Cosmology and Fundamental Physics \linebreak
  $\square$  Stars and Stellar Evolution \hspace*{1pt} $\square$ Resolved Stellar Populations and their Environments \hspace*{40pt} \linebreak
  $\square$    Galaxy Evolution   \hspace*{45pt} $\square$             Multi-Messenger Astronomy and Astrophysics \hspace*{65pt} \linebreak
  
\noindent \textbf{Principal Author: }

\noindent Name:	Diana Dragomir \\
Institution:  MIT/University of New Mexico \\
Email:  dragomir@unm.edu \\
Phone:  617-324-1324 \\
 
\noindent \textbf{Co-authors:} Eliza Kempton (University of Maryland), Jacob Bean (University of Chicago), Ian Crossfield (MIT), Eric Gaidos (University of Hawaii), Nikole Lewis (Cornell University), Michael Line (Arizona State University), Roxana Lupu (BAER Institute/NASA Ames), George Zhou (Harvard CfA)

\noindent \textbf{Co-signers:} Daniel Angerhausen (Universitat Bern), Jasmina Blecic (NYUAD), David R. Ciardi (Caltech-NExScI), Jonathan Fortney (UC Santa Cruz), Nicolas Iro (University of Vienna), Marshall C. Johnson (The Ohio State University), Quinn Konopacky (UC San Diego), Dimitri Mawet (Caltech), Henry Ngo (National Research Council Canada), Chris Packham (University of Texas at San Antonio), Seth Redfield (Wesleyan University), Tyler Robinson (Northern Arizona University), Gautam Vasisht (JPL), Jason T. Wright (The Pennsylvania State University) \\
 
\noindent \textbf{Abstract:} 
The best-characterized exoplanets to date are planets on close-in transiting orbits around their host stars. The high level of irradiation and transiting geometry of these objects make them ideal targets for atmospheric investigations. However, the modest apertures of many current telescopes allow mostly low resolution spectra to be observed for transiting planets, failing to extract key physical and chemical properties of their atmospheres. Ground-based 30-meter class telescopes will set the stage for a substantial leap in our understanding of exoplanet atmospheres. We outline a two-pronged survey, recently submitted as a Key Science Project (KSP) for the US ELTs, which would yield unprecedented insight into the atmospheres of close-in exoplanets via combined observations with the GMT and TMT.  (1) The first opportunity involves measuring the global-scale atmospheric circulation and planetary rotation for a sample of 40 hot Jupiters to glean insight into the unique radiative forcing regime governing highly-irradiated, tidally-locked giant planets.  (2) The second opportunity involves measuring  atmospheric mass-loss and extracting atmospheric composition and abundance ratio information for $\sim60$ sub-Neptunes and super-Earths (including candidate disintegrating planets) to constrain their formation and evolution histories.  These efforts would be made possible by the unparalleled combination of high spectral resolution instrumentation and large aperture size of the ELTs.  This survey would enable the first statistical study of atmospheric circulation in extrasolar giant planets, and would provide detections of trace gases and measurements of atmospheric escape in small-planet atmospheres, far exceeding the reach of \textit{JWST}.

\pagebreak

\subsection*{A Unique Window into Exoplanet Atmospheres Using High Resolution Spectroscopy}

\begin{wrapfigure}{R}{0.5\textwidth} 
\vspace{-20pt}
\centering
\includegraphics[width=8cm]{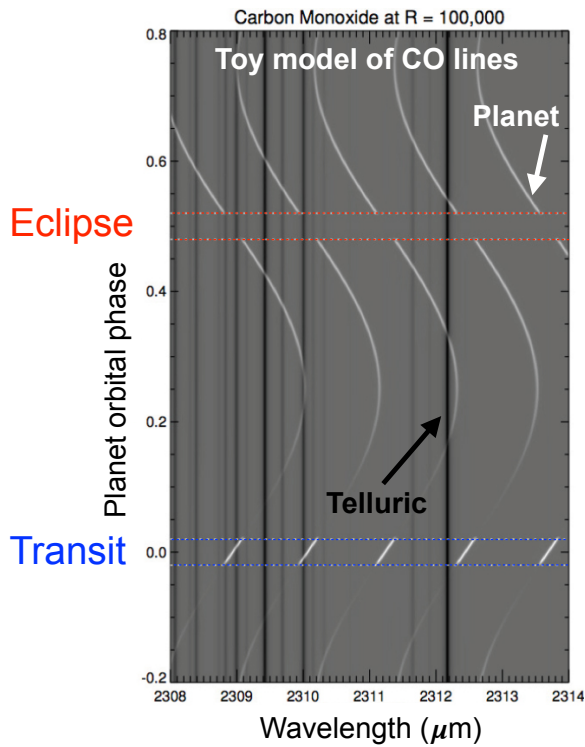}
\caption{\label{fig:COtoy} ``Toy" model demonstrating how lines in an exoplanet spectrum are separated from stellar lines and tellurics by the orbital motion of the planet (typically $\sim$100 km~s$^{-1}$).  The transmission spectrum through the planet's upper atmosphere is obtained during transit.  At all other orbital phases except secondary eclipse, the planet's thermal emission is obtained. (Adapted from figure by Matteo Brogi.)}
\vspace{-15pt}
\end{wrapfigure}

Measurements of the composition and dynamics of a planet's atmosphere can, in principle, reveal information about its origin and evolution. In practice, attempts to achieve this have so far largely been stymied by limitations of the instruments and techniques used. The most common technique used to probe exoplanet atmospheres is low-resolution transmission spectroscopy, which measures the apparent size of a transiting exoplanet as a function of wavelength. Though it has had some successes \citep{Fra14,Wak17}, low-resolution transmission spectroscopy has more often than not been foiled by the presence of aerosols, which appear to be prevalent in small exoplanet atmospheres \citep{Kre14a, Knu14, Knu14b, Ehr14} but can also be found in hot Jupiter atmospheres \citep{Sin16}. 

For transiting systems (as well as non-transiting close-in planets), an observing technique that has shown great promise and offers key advantages over low-resolution spectroscopy is high-resolution cross-correlation spectroscopy (HRCCS). High-resolution spectroscopy is already established as a powerful technique for exoplanet mass and obliquity measurements (see companion white papers by Johnson et al. and Ciardi et al.), but these measurements are based on the stellar spectral lines. \textbf{HRCCS goes one step further: it allows for exoplanetary absorption or emission lines to be spectrally resolved from both the stellar spectrum and telluric absorption lines}, owing to the orbital motion of the planet around its host star (Figure 1). The use of cross correlation techniques against model templates then allows for the spectral signature of the molecules in the exoplanet's atmosphere to be extracted, even in instances when the planet's spectrum is only sampled at very low signal-to-noise (S/N). 

In this white paper, we describe how the advent of the Extremely Large Telescopes (ELTs) can tremendously expand the science potential of the HRCCS technique to observe the atmospheres of close-in exoplanets from new perspectives. For giant planets, the technique makes possible direct measurements of rotation and atmospheric circulation at unprecedented precision in order to better understand how radiative forcing determines the three-dimensional (3-D) structure of close-in giant planet atmospheres. For small planets, HRCCS measurements can constrain the composition of their atmospheres and crusts, as well as their rates of atmospheric escape. These constraints will provide insight into the formation history, interior composition, and atmospheric evolution of small planet atmospheres that were not necessarily generated through accretion of nebular gas.

\subsection*{Dynamics and Rotation of Giant Planet Atmospheres}

\begin{wrapfigure}{R}{0.53\textwidth} 

\centering
\includegraphics[width=8cm]{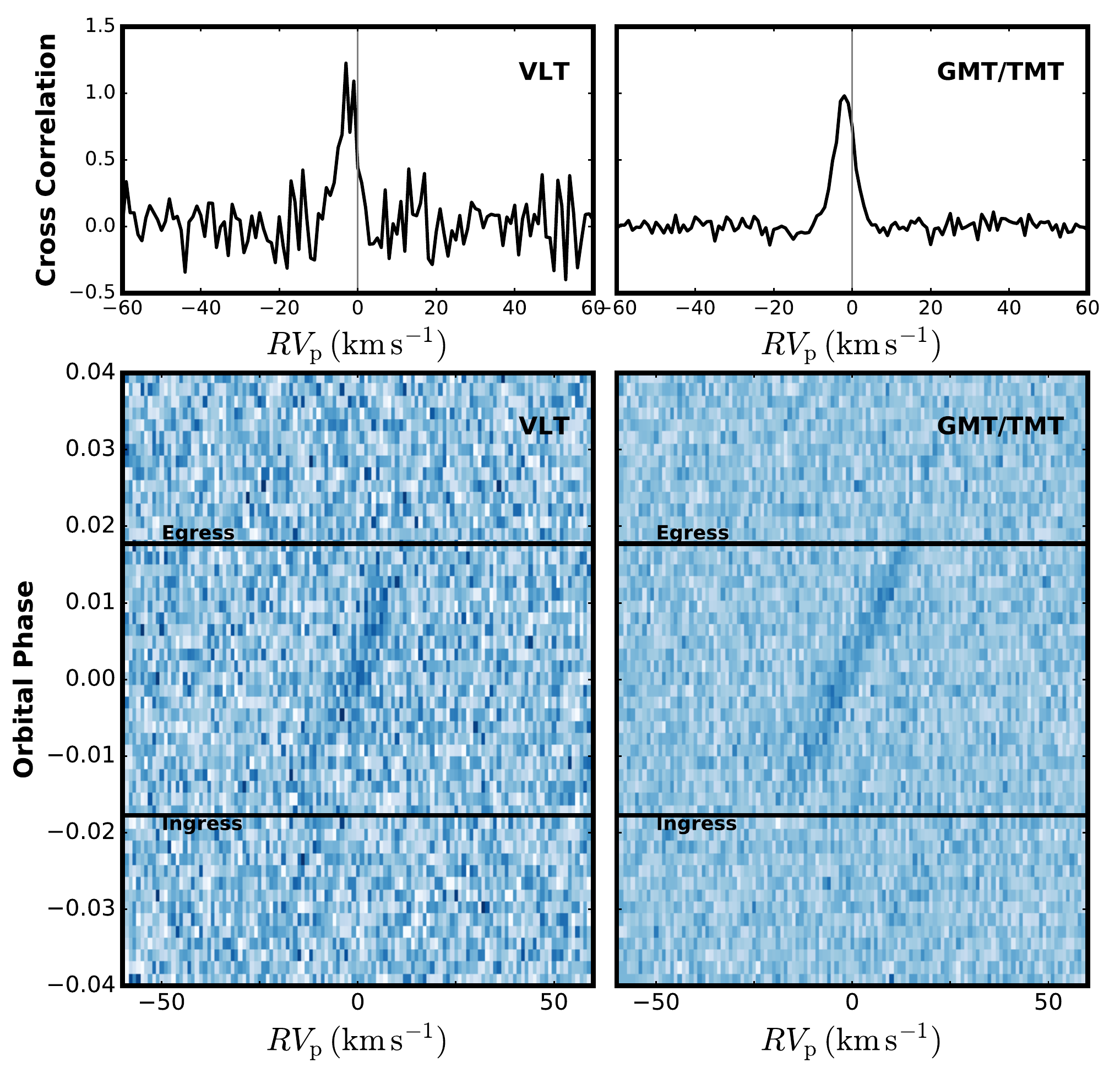}
\caption{\label{fig:CrossCor} Simulated cross correlation results for one transit of the hot Jupiter HD 209458b with CRIRES on the VLT (\citealt{Sne10}; left), compared to expected K-band observations with the US ELTs (right). The dark diagonal band in the lower panels indicates the planet's orbital motion.  Day-to-night winds with magnitude 2 km s$^{-1}$ are shown by the offset of the cross-correlation signal (already corrected for orbital motion) from $v = 0$, which is apparent in the upper panels. We expect a $\sim$fourfold improvement in precision on constraining the exoplanetary velocity field for the US ELTs over current facilities.}
\vspace{-10pt}
\end{wrapfigure}

The unique opportunity offered up by HRCCS for giant planet atmospheres is to directly probe their winds and rotation through Doppler shifted and broadened spectral lines.  The main motivation for observing these signatures of atmospheric dynamics is to understand energy transport processes in irradiated planetary atmospheres.  Hot Jupiters are expected to be tidally locked with hot permanent daysides, which places them in a particularly interesting radiative forcing regime. 3-D models of hot Jupiter atmospheres almost uniformly predict super-rotating equatorial jets and an eastward-shifted global hot spot \citep[e.g.][]{sho09,rau10}, but key effects are typically omitted --- notably magnetic fields, disequilibrium chemistry, and shocks.  Observations that directly probe atmospheric circulation are therefore needed to guide the development of theory and to establish which physical effects are important.

In a typical hot Jupiter, both the expected wind speeds and equatorial rotation velocities have values $\sim$1-10 km s$^{-1}$, so spectral resolutions approaching $10^5$ are needed to detect their effects.  In transmission spectra, day-to-night winds at $\sim$mbar pressures in hot Jupiters directed toward the observer are expected to induce $\sim$2 km s$^{-1}$ net blue-shifts \citep{kem12, sho13}.  In thermal emission, the combination of strong equatorial super-rotating jets, tidally locked rotation, and the emission signature primarily originating near the planet's dayside equator, should produce net Doppler shifts of $\sim$1-10 km~s$^{-1}$ \citep{zha17}.  Current facilities have so far yielded only tentative measurements of atmospheric dynamics for a small number of the most favorable targets \citep[e.g.][]{Sne10, Bro16}.  \textbf{The much larger apertures of the ELTs are needed for robust detection of atmospheric circulation (Figure 2) and to expand the reach of these studies beyond the few very brightest systems into a regime in which statistical investigations are possible.} Specifically, $\lambda/\Delta \lambda \sim 10^5$ instruments on the ELTs could measure the atmospheric dynamics for 40 hot Jupiters, half of which are known and half of which will be discovered by the \emph{TESS} (Transiting Exoplanet Survey Satellite) mission (as estimated from the simulated planet yields of \citealt{Sul15} and \citealt{Hua18}).

Additional dynamical effects that can be probed through HRCCS include planetary rotation, and vertical and longitudinal wind gradients \citep{snellen14, kem12}.  Planetary rotation typically broadens spectral lines in transmission, while the effects of rotation are more complex in thermal emission spectra because of the dominance of the hot spot in the emitted flux.  Measurements of rotation rates via the shape and width of the obtained cross correlation function can be used to test the prediction that the orbits of close-in planets are tidally synchronized.  Rotation, as well as longitudinal wind gradients, can also be probed through differential ingress-egress transmission spectroscopy by separately measuring the absorption through the planet's eastern and western limbs, or through thermal emission phase curves with high resolution spectra obtained over a broad range of orbital phases. Vertical wind shear can be assessed through measuring the Doppler shifts of individual absorption lines in transmission spectra. The strongest lines probe the lowest pressures in the planetary atmosphere, allowing the line-of-sight wind speeds to be directly measured as a function of altitude.  \textbf{Both ingress-egress comparison spectroscopy and measurements of vertical wind shear are not possible from current facilities, but higher S/N spectra obtainable with the ELTs will bring these effects within our observational reach.}

In addition to probing atmospheric dynamics, HRCCS observations can be used to set quantitative constraints on abundances and temperatures of hot Jupiters \citep{bro18}, even though they suffer from the loss of the continuum and absolute flux levels due to the telluric/stellar removal methods. Indeed, information about the atmospheric physics maps through the planetary line ratios and line-to-continuum contrast via the cross-correlation coefficient. This can in turn be mapped into a log-likelihood function that enables typical Bayesian retrieval analyses \citep[e.g.][]{Lin13, mad18}. The larger aperture of the ELTs will also open the door to exploring other subtle diagnostics that are inaccessible with current facilities.  An example is the potential to probe isotopic ratios in exoplanetary spectra \citep{mol18}, which also necessitates high spectral resolution and high S/N.  

\subsection*{Small Planet Atmospheres: Atmospheric Composition and Escape}

While HRCCS of close-in giant exoplanets is already expanding beyond detection of molecules into the realm of atmospheric dynamics, we are just beginning to scratch the surface of this technique's potential for probing the atmospheres of super-Earths and sub-Neptunes. Efforts to probe their atmospheres with low-resolution transmission spectroscopy have often revealed the presence of clouds or hazes that obscure spectral features \citep{Kre14a,Knu14,Knu14b}. This limitation suggests an obvious application of HRCCS: by probing at higher altitudes than low-resolution spectroscopy, \textbf{HRCCS offers the opportunity to observe above the cloud deck and determine atmospheric composition and abundance ratios \citep{bro18} even for aerosol-dominated planets.} But the prospects of this technique for small planets and their atmospheres go far beyond simply ``seeing above the clouds".

Close-in Neptune-sized exoplanets are believed to have retained their primordial atmospheres. In addition, the recent ground-based detection of He in the atmosphere of the exo-Neptune HAT-P-11b \citep{All18} with the CARMENES instrument (R$\sim$80,000) has revealed an important means of studying atmospheric escape on these planets. While they are powerful facilities, space-based telescopes like \textit{HST}, \textit{Spitzer}, and \textit{JWST} are equipped with low-resolution spectrographs that do not resolve the metastable He line sufficiently to enable precise constraints on atmospheric escape rates. Upgrading HRCCS to a ELT will allow us to use measurements of He and other atomic species to constrain atmospheric loss for dozens more Neptune-sized exoplanets (most of which will be discovered by \emph{TESS}; \citealt{Ric15}). Directly measuring the atmospheric composition of close-in Neptune-sized planets can also trace their evolution. High atmospheric mass-loss rates would indicate that these planets could be on their way to crossing the photo-evaporation radius gap \citep{Ful17}, suggesting that they will eventually become rocky cores stripped of their primordial atmospheres. The high spectral resolution and high S/N afforded by the ELTs will furthermore allow for the thermal and velocity structure of the escaping gas to be probed. Together with constraints on the stellar age, mass-loss measurements can help determine how much more massive the planets were when they formed, thus testing whether they formed with close to their current envelope of gas, or whether they are remnants of evaporated hot Jupiters. These constraints on the planetary histories are complementary to those that can be gained with orbital obliquity measurements (see white paper by Johnson et al.).

Conversely, hot super-Earths ($R_p \lesssim 2 R_{\oplus}$) are not believed to be sufficiently massive to retain the H/He-dominated atmospheres they may have been born with. Instead, highly-irradiated rocky super-Earths likely have secondary atmospheres that are outgassed, or composed of silicate vapors originating from the planets' mantles. To date, only one super-Earth has been observed with the HRCCS technique: 55 Cnc e. \citep{Rid16} reported tentative detections of Na and Ca$^+$ via five transits of 55 Cnc e with several current large-aperture telescopes. With an instrument such as G-CLEF on GMT, we would be able to both verify this result and detect other species with a single transit of 55 Cnc e. The heavy secondary atmospheres of super-Earths are not expected to escape as easily as lighter primordial atmospheres, but even so, a number of species are expected at the low pressures probed by HRCCS. Moreover, when observed out of transit, hot super-Earths (either rocky or volatile-rich) make excellent targets for emission spectroscopy. Figure 3 shows a simulated spectrum of the flux emitted by a hot rocky planet, which includes absorption features from molecules and atoms such as H$_2$O, CO$_2$, OH, Na and K. As a result, at high spectral resolution, features due to these species as well as less common gases such as O, O$_2$, SiO SO$_2$, H$_2$S, HCl, HF, NaCl, and KCl should be present in transmission and/or emission spectroscopy \citep{Sch12}. These species have spectral features across the optical and IR wavelength range that will be accessible to at least some of the instruments planned for the ELTs. Furthermore, the atmospheric pressures probed with HRCCS span a wider dynamic range when compared with low-resolution space-based spectroscopy, therefore also allowing for stronger constraints on thermal structures.  

From simulations of the expected planet yield from \emph{TESS} \citep{Sul15, Hua18}, \textbf{we estimate that 50 hot sub-Neptunes, super-Earths and Earth-sized planets will be discovered transiting sufficiently nearby stars such that their atmospheres will be accessible to HRCCS observations with the ELTs.}

\begin{figure}
\begin{center}
\includegraphics[width=16.5cm]{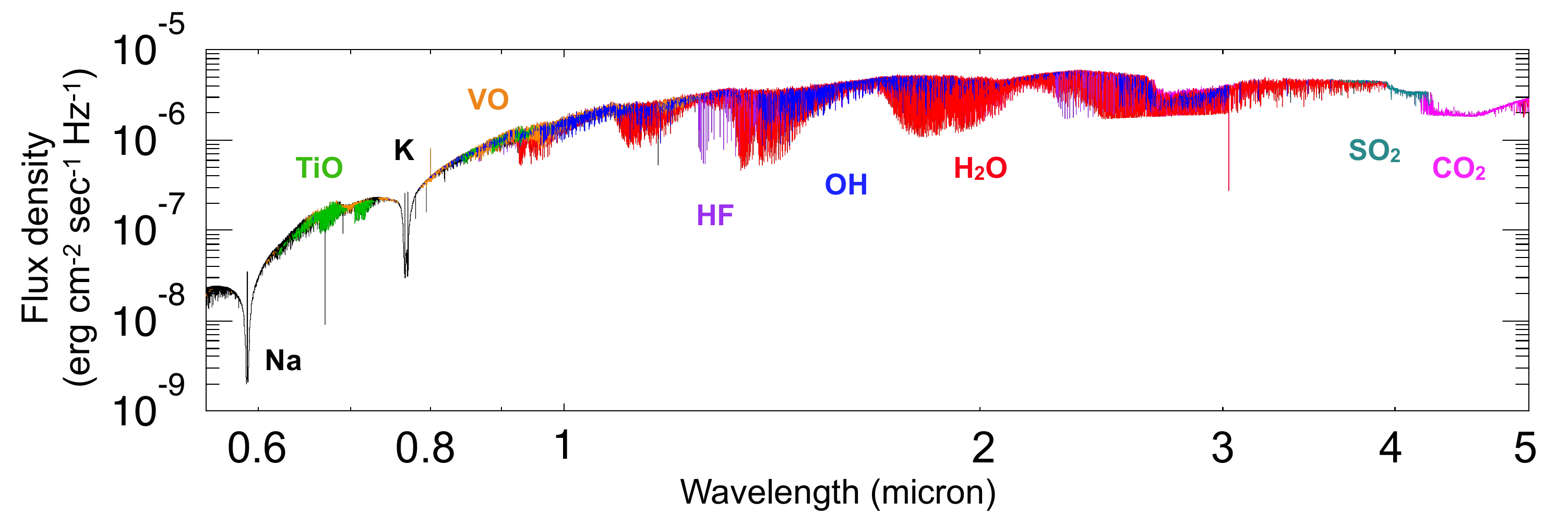}
\caption{\label{fig:hotcrust} Simulated emission spectrum for a hot rocky planet, with an atmosphere resulting from the evaporation of rock material with the composition of the Earth's continental crust. The planet is situated at 0.01 AU from its host star.} 
\end{center}
\vspace{-20pt}
\end{figure}

Finally, a few ``disintegrating" planets have been discovered. Too small for their radius to be measured, these are likely Mercury-size rocky planets on very short-period orbits which have presumably lost their volatiles and on which star-ward surface temperature ($\gtrsim 2000$K) produce a silicate (``rock") vapor atmosphere that entrains dust and forms a cometary tail that is detectable ($\sim$1\%) in transit \citep{Rap12,per13,San15}. These represent an opportunity to probe the surface and interior of rocky cores, via the emissivity properties of the mineral grains \citep{Bod18} or by investigations of the vapor component (volatile elements such as sodium and refractory elements such as Ca, Al, and Si) in transmission.  Observations of the latter require high spectral resolution, high signal-to-noise, and/or rapid cadence of comparatively faint host stars, thus requiring the large apertures of ELTs. The all-sky \emph{TESS} mission is expected to yield $\sim$10 additional examples of such objects around brighter stars (Gaidos et al., in prep.). The combination of both US ELTs (TMT and GMT) would provide sky access for the full set of candidates, plus greater longitudinal coverage for uninterrupted observations of the transit of the dust cloud, which can be significantly larger in extent than the host planet. 

In total, by constraining the atmospheric composition, thermal structure, and escape rate of hot super-Earths and sub-Neptunes through HRCCS with the ELTs, we will shed new light on their formation, evolution, dynamical history, and the connection between core and stellar composition.

\subsection*{Conclusion and Recommendations}

We wish to echo the findings and recommendations of the National Academy of Sciences \textit{Exoplanet Science Strategy Consensus Study Report} --- specifically the recommendation for investment in ELTs.  Ground-based ELTs have unique capabilities to bring to the characterization of close-in exoplanets.  Chief among these is the high-resolution spectroscopic characterization of exoplanet atmospheres, which is the most promising technique to directly measure atmospheric winds and constrain abundances of trace gases, even in cloudy or hazy atmospheres.  At present, the European ELT is the only fully funded facility that will be able to measure the exoplanet properties detailed in this white paper.  The two US ELT facilities would allow the American astronomical community to have a leadership role in ground-based characterization of exoplanet atmospheres into the next decade and beyond.  Furthermore, the combination of both the GMT and TMT will provide full-sky access and sufficient observing time to conduct surveys of $\sim$10s of objects to search for statistical trends.  For these reasons we recommend that the decadal survey committee support the two US ELT projects. \\

\bibliographystyle{aasjournal_DD}
\bibliography{research}

\end{document}